\documentclass{PoS}

\usepackage{amsfonts}
\usepackage{amsmath}
\usepackage{subfigure} % Include figure files

\graphicspath{{./figs/}}
\title{
  $B_K$ with improved staggered fermions: analysis using
SU(3) staggered chiral perturbation theory 
}

\ShortTitle{
  SU(3) SChPT analysis of $B_K$
}

\author{\speaker{Jangho Kim}, Taegil Bae, Yong-Chull Jang,
  Hyung-Jin Kim, Jongjeong Kim, Kwangwoo Kim, Boram Yoon, Weonjong Lee\\
  Lattice Gauge Theory Research Center, CTP, and FPRD, \\
  Department of Physics and Astronomy,
  Seoul National University, Seoul, 151-747, South Korea \\
  E-mail: \email{wlee@snu.ac.kr}}

\author{Chulwoo Jung \\
  Physics Department, Brookhaven National Laboratory,
  Upton, NY11973, USA \\
  E-mail: \email{chulwoo@bnl.gov}}

\author{Stephen R. Sharpe\\
  Physics Department, University of Washington, Seattle, WA 98195-1560 \\
  E-mail: \email{sharpe@phys.washington.edu}}

\abstract{ We report updated results for $B_K$ using HYP-smeared
 staggered valence quarks on MILC asqtad lattices based on an analysis
using SU(3) staggered chiral perturbation theory.  
The most important new feature of our data sample is the
inclusion of a fourth (``ultrafine'') lattice spacing.
This improves the control over the continuum extrapolation and
errors due our use of one-loop perturbative matching.
We present a complete updated error budget, which leads to
$B_K(\text{NDR}, \mu = 2 \text{ GeV}) = 0.5309 \pm 0.0051 \pm  0.0424$ 
and 
$\hat{B}_K = B_K(\text{RGI}) = 0.727 \pm 0.07 \pm 0.058$. 
The results of the SU(3) analysis are inferior to those based on SU(2)
staggered chiral perturbation theory, primarily because of
the dependence on the Bayesian priors we use in the SU(3) fits.
}

\FullConference{
  The XXVIII International Symposium on Lattice Field Theory, Lattice2010\\
  June 14-19, 2010\\
  Villasimius, Italy
}

\begin{document}

\section{Introduction} 
This paper is the second in a series of four providing
an update on 
our determination of $B_K$ using improved staggered fermions.
Here, we update the results obtained using fit functions
derived from SU(3) staggered chiral perturbation theory (SChPT).
In particular, we focus on the progress since last year's
lattice proceedings~\cite{ref:wlee-2009-1}.
Our present results are based on 10 ensembles of MILC asqtad lattices,
whose properties are listed in Table~\ref{tab:milc-lat}.
As shown in the table, in the last year we have 
substantially increased the number of measurements on two ensembles,
and added two new ensembles.
In particular, the addition of the ultrafine ensemble U1 means
that we now have 4 lattice spacings, as opposed to 3 last year.

During the last year, we have also prepared a long article in which
we give a detailed description of all aspects of the calculation,
and present final results based on 3 lattice spacings~\cite{ref:wlee-2010-1}.
This reference contains all the details that we must necessarily skim over
here due to space constraints.
We also adopt the notation of Ref.~\cite{ref:wlee-2010-1}
for parameters and fit types, and warn the reader that the labels for fits
are different from those used in Ref.~\cite{ref:wlee-2009-1}.
\begin{table}[h!]
\begin{center}
\begin{tabular}{c | c | c | c | c | c | c}
\hline
$a$ (fm) & $am_\ell/am_s$ & geometry & ID & ens $\times$ meas 
& $B_K$ (N-BB1) & $B_K$ (N-BB2) \\
\hline
0.12 & 0.03/0.05  & $20^3 \times 64$ & C1 & $564 \times 1$ &  0.555(12) & 0.564(17) \\
0.12 & 0.02/0.05  & $20^3 \times 64$ & C2 & $486 \times 1$ &  0.538(12) & 0.535(17) \\
0.12 & 0.01/0.05  & $20^3 \times 64$ & C3 & $671 \times 9$ &  0.562(6)  & 0.592(14) \\
0.12 & 0.01/0.05  & $28^3 \times 64$ & C3-2 & $275 \times 8$ &  0.575(6)  & 0.595(13) \\
0.12 & 0.007/0.05 & $20^3 \times 64$ & C4${}^*$ & $651 \times 10$ & 0.564(5)  & 0.598(13) \\
0.12 & 0.005/0.05 & $24^3 \times 64$ & C5 & $509 \times 1$ &  0.567(10) & 0.588(19) \\
\hline
0.09 & 0.0062/0.031 & $28^3 \times 96$ & F1 & $995 \times 1$ & 0.535(9) & 0.539(12) \\
0.09 & 0.0031/0.031 & $40^3 \times 96$ & F2${}^{\#}$ & $678 \times 1$ & 0.540(8) & 0.545(13)\\
\hline
0.06 & 0.0036/0.018 & $48^3 \times 144$ & S1${}^*$ & $744 \times 2$ & 0.535(6) & 0.560(11)\\
\hline
0.045 & 0.0028/0.014 & $64^3 \times 192$ & U1${}^{\#}$ & $305 \times 1$ & 0.540(6) & 0.547(8)\\
\hline
\end{tabular}
\end{center}
\caption{MILC asqtad ensembles used in the calculation.  Ensembles
marked with a ${}^*$ have improved statistics compared to last year,
while those marked with a ${}^\#$ are new. Results for $B_K(\mu=2\
{\rm GeV})$ using both N-BB1 and N-BB2 fits are given. See text for
discussion of these fits.}
\label{tab:milc-lat}
\end{table}

\section{SU(3) SChPT Analysis}
SU(3) SChPT was developed in Refs.~\cite{ref:wlee-1999-1,ref:bernard-2003-1}
and first applied to a calculation of $B_K$ in Ref.~\cite{ref:sharpe-2006-1}.
Since we use a mixed action, we need to generalize the SChPT calculation, and
have done so in Ref.~\cite{ref:wlee-2010-1}.
The result is that, for fixed $a$ and sea-quark masses,
the next-to-leading (NLO) order expression contains 14 low-energy coefficients
(LECs).
To obtain a good fit we also need to add a single analytic NNLO term,
so that there are 15 LECs in all.
Of these, 11 are due to lattice artifacts---either discretization errors or
errors due to our truncation of the matching factors at one-loop order.
The other 4 LECs remain in the continuum limit.

On each ensemble we have 10 valence quark masses
(running from $\sim m_s^{\rm phys}$ down to $\sim m_s^{\rm phys}/10$)
and thus 55 different kaons.
Nevertheless, a direct fit using all 15 parameters is not stable, 
primarily because several fit functions are similar.
To proceed, we reduce the number of parameters to 7 by removing 8
lattice artifact terms which have similar functional dependence to
one of the 3 such terms that we keep.
The details of this procedure are described in
Ref.~\cite{ref:wlee-2010-1}.
This then allows stable fits.
However, we find that, in some of these fits, the coefficients
of the remaining lattice artifact terms come out larger than
one would expect based on naive dimensional analysis.
Thus as a second modification we 
constrain the size of the coefficients of these coefficients.
We try different schemes for these constraints,
as explained in Ref.~\cite{ref:wlee-2010-1}.

We focus here on what we consider our most reliable SU(3)-based approach.
This is based on a two step fitting procedure, in which we first
fit only to degenerate points ($m_x=m_y$), and then
use the results of this fit as constraints on parameters for a fit
to the entire data set.
This approach makes sense because the
the fitting form is much simpler for the degenerate kaons,
containing only 4 parameters (3 continuum and 1 lattice artifact).
Fits to our 10 data points are stable and do not require Bayesian
constraints, although we include such constraints for consistency,
as we do need them in the second stage of the fitting.

In slightly more detail, we fit to the form
\begin{eqnarray}
f_\text{th}^\text{deg} &=& \sum_{i=1}^{4} c_i F_i
\label{eq:fit:deg}
%\\
%F_i &=& H_i \qquad \text{for }\ \ i=1,2,3
%\\
%F_4 &=& H_{10} = F^{(4)}_T
\end{eqnarray}
where the functions $F_i$ depend on the pion and kaon masses
and are given in Ref.~\cite{ref:wlee-2010-1}.
$F_4$ is the  lattice artifact term, and we
constrain the LEC $c_4$ by augmenting the $\chi^2$:
\begin{equation}
\chi^2_\text{aug} = \chi^2 + \chi^2_\text{prior}\,; \qquad
\chi^2_\text{prior} = {(c_4 - a_4)^2}/{\tilde\sigma_4^2}
\end{equation}
We set $a_4 = 0$ (since we do not have prior knowledge of the sign
of $c_4$), and use
$\tilde\sigma_4 \approx \Lambda_\text{QCD}^2 (a \Lambda_\text{QCD})^2$
for the ``D-B1'' fit or
$\tilde\sigma_4 \approx \Lambda_\text{QCD}^2 \alpha_s^2$
for the ``D-B2'' fit.
These two choices assume, respectively, that $c_4$ is dominated by either
discretization errors or truncation errors in matching.
As noted above, these constraints have little impact on the
degenerate fits.

In the second stage, we extend the fit to the full data set,
using the fit function
\begin{eqnarray}
f_\text{th}^\text{non-deg} &=& \sum_{i=1}^{7} c_i F_i
%\\
%F_5 &=& H_4 \\
%F_6 &=& H_{12} = F^{(6)} \\
%F_7 &=& H_{14} = F^{(1)}_A
\end{eqnarray}
where the first 4 terms are the same as in Eq.~(\ref{eq:fit:deg}) and
the rest are given in Ref.~\cite{ref:wlee-2010-1}.
Of the 3 new terms, only $F_5$ survives in the continuum limit.
We augment the $\chi^2$ with
\begin{eqnarray}
\chi^2_\text{prior} &=& 
\chi^2_\text{prior (1)} + \chi^2_\text{prior (2)}
\\
\chi^2_\text{prior (1)} &=& 
\sum_{i=1}^{4} {(c_i - a_i)^2}/{\tilde\sigma_i^2}\,,\quad
\chi^2_\text{prior (2)} = 
\sum_{j=6,7} {c_j^2}/{ \tilde\sigma_j^2}
\end{eqnarray}
Here $a_i \pm \tilde\sigma_i$ are the results of either the D-B1 or D-B2 fit,
which feeds in the information from the degenerate fits.
%
%Note that the fitting results for the D-B1 (D-B2) fit becomes the
%prior information for the N-BB1 (N-BB2).
%
The other priors are
\begin{eqnarray}
\tilde\sigma_6 &=& \left\{\begin{array}{l l} 
    \Lambda_\text{QCD}^2 (a \Lambda_\text{QCD})^2  &  \text{for N-BB1 fit} \\
    \Lambda_\text{QCD}^2 \alpha_s^2                & \text{for N-BB2 fit} 
\end{array}\right. \,,
\\
\tilde\sigma_7 &=& \left\{\begin{array}{l l} 
    \Lambda_\text{QCD}^4 (a \Lambda_\text{QCD})^2  &  \text{for N-BB1 fit} \\
    \Lambda_\text{QCD}^4 \alpha_s^2                & \text{for N-BB2 fit} 
\end{array}\right. \,.
\end{eqnarray} 
These constrain the lattice artifact terms and have a significant impact
on the fits.

\section{Fitting and Results}
%
%
%
%--------------------
% C4 and F2 lattices
%--------------------
\begin{figure}[t!]
\centering
\includegraphics[width=0.49\textwidth]
{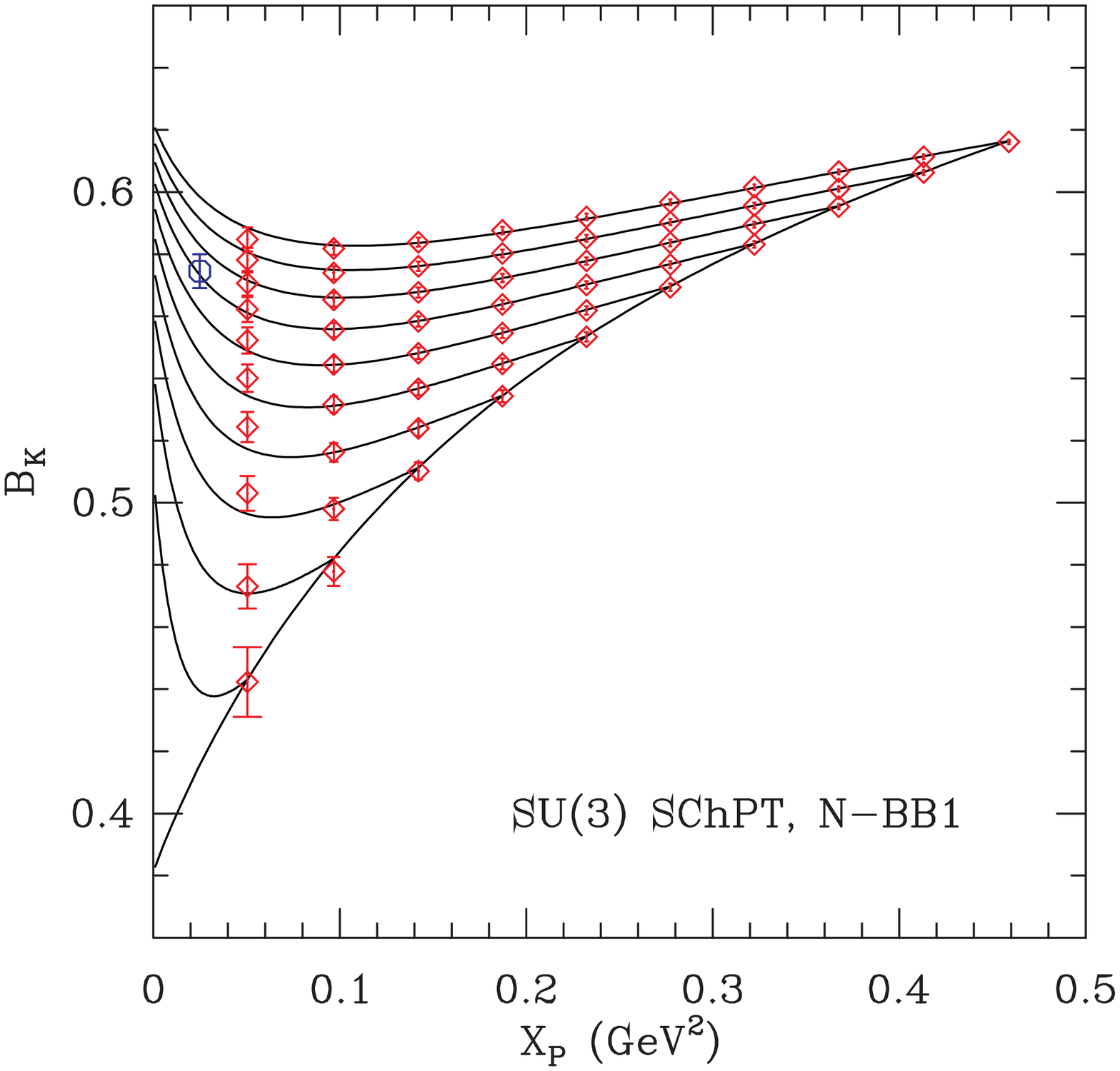}
\includegraphics[width=0.49\textwidth]
{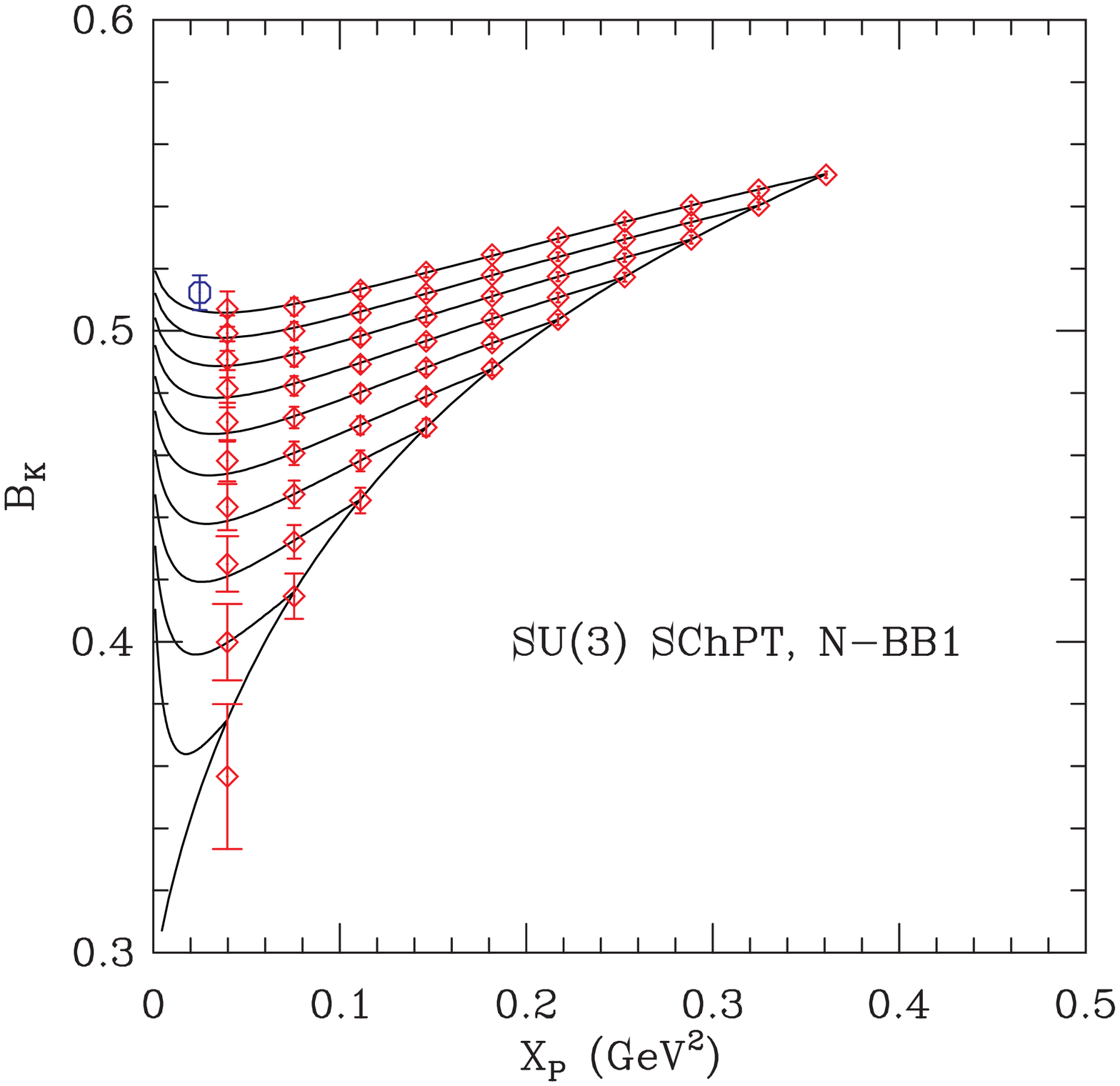}
\caption{$B_K$ (obtained using one-loop matching) versus $X_P$ (squared
mass of pion composed of valence $d$ and $\bar d$) for the C4 (left)
  and S1 (right) ensembles. N-BB1 fits are shown.
  The red diamonds show the raw data, while
  the blue octagon shows the result obtained after extrapolation to
 to the physical quark masses with all taste-breaking lattice artifacts removed.}
\label{fig:su3,N-BB1,C4+S1}
\end{figure}
In Fig~\ref{fig:su3,N-BB1,C4+S1}, we show results of the N-BB1 fit on
the C4 and S1 ensembles.
Compared to last year~\cite{ref:wlee-2009-1},
we have increased the statistics by factors of 10 (C4) and nearly 3 (S1).
We also plot the data in a new way, showing it as a function of the
squared mass of the pion composed of two light valence quarks.
This displays the extrapolation to the physical kaon more clearly.
Despite the reduction in error bars compared to last year
(particularly in the left panel), the
fit form gives a reasonable representation of the data.
In Fig~\ref{fig:su3,N-BB1,F1+F2}, we show a similar plot 
for the F1 and F2 ensembles, the latter being new.
Again, the fits are reasonable.
The quality of the N-BB2 fits is similar on all ensembles.
Fits for the U1 ensemble are described in a companion 
proceedings~\cite{ref:wlee-2010-5}.

\begin{figure}[t!]
\centering
\includegraphics[width=0.49\textwidth]
{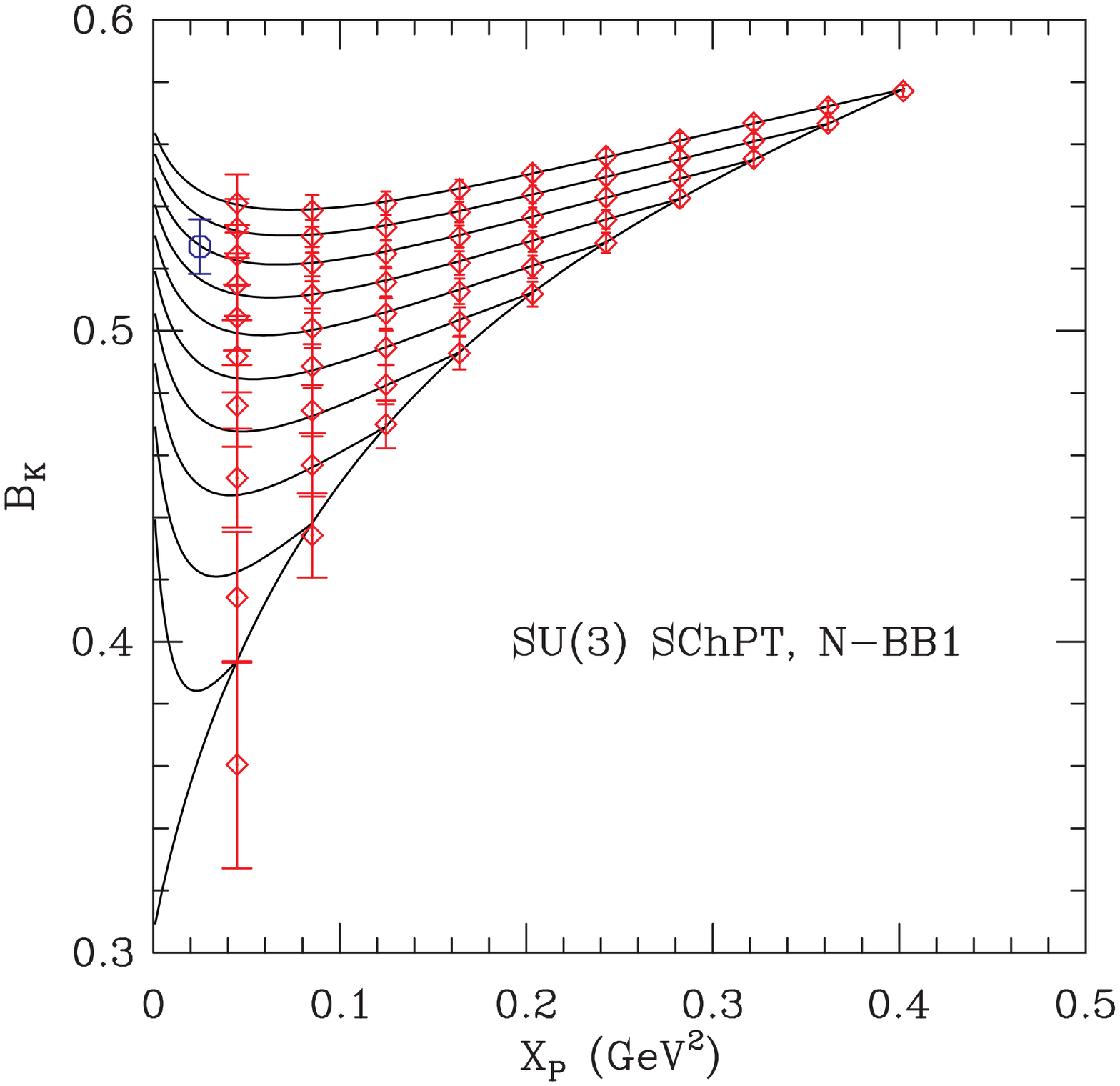}
\includegraphics[width=0.49\textwidth]
{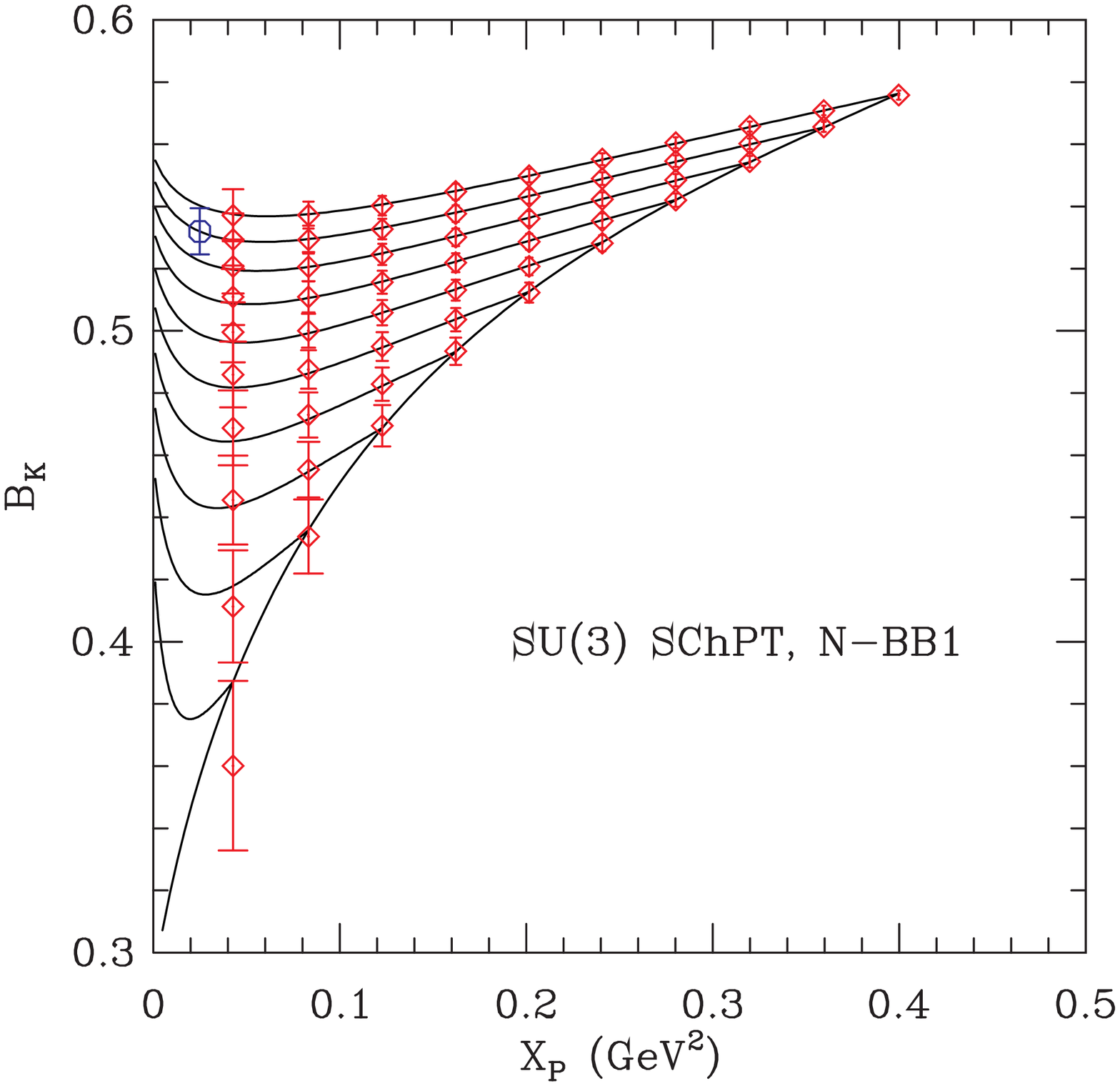}
\caption{As in Fig.~\protect\ref{fig:su3,N-BB1,C4+S1} but for
  the F1 (left) and F2 (right) ensembles.
}
\label{fig:su3,N-BB1,F1+F2}
\end{figure}

Results for $B_K$ from
both types of fit are given in Table~\ref{tab:milc-lat}. 
We find that, on each ensemble, 
the two fits are consistent within 2$\sigma$,
but that $\sigma$ is relatively large,
so the central values differ be as much as $\sim 5\%$.
%(on ensembles C3, C4 and S1).
%
This indicates a significant sensitivity to the
size of the terms representing lattice artifacts,
whose values cannot be pinned down by our fits alone.
%
%This is the main drawback of the SU(3)
%analysis compared with that based on SU(2) SChPT.
%

A striking result of the fits is that the coefficient $c_5$ is
very small. This multiplies
\begin{equation}
F_5 \propto (m_d^{\rm val}-m_s^{\rm val})^2/(m_d^{\rm val}+m_s^{\rm val})
\,,
\end{equation}
which is the sole continuum term contributing only
for non-degenerate kaons. The expectation is that $c_5$
should be of $O(1)$, but it appears to be more than an order
of magnitude smaller. This result implies that, in the continuum limit,
we can almost determine $B_K$ at the physical, non-degenerate point, 
using only degenerate kaons. This gives further {\em a posteriori}
justification to our two-stage fitting procedure.

A concern with these fits is poor convergence of
SChPT. One can see from the figures that the result in the
chiral limit (obtained by extrapolating the degenerate points to
$X_P=0$), which is the LO term in ChPT, lies substantially below
the final extrapolated result for $B_K$ (given by the blue octagons).
For more detailed discussion, see Ref.~\cite{ref:wlee-2010-1}.

\section{Continuum Extrapolation}
We use the results from the C3, F1, S1 and U1 ensembles to do
the continuum extrapolation, since all have approximately the
same values for $m_\ell$ and $m_s$ (and, in particular, the
same ratio $m_\ell/m_s$).
We then extrapolate to the physical values of these masses,
based on the dependence seen on the coarse and fine lattices.
Note that the non-analytic part of this dependence has already been taken into
account by setting $m_\ell=m_\ell^{\rm phys}$
and $m_s=m_s^{\rm phys}$ in the SChPT fit forms.
In fact, the remaining dependence on $m_\ell$ is very weak, as can
be seen from Table~\ref{tab:milc-lat}.

\begin{figure}[t!]
\centering
\includegraphics[width=0.49\textwidth]{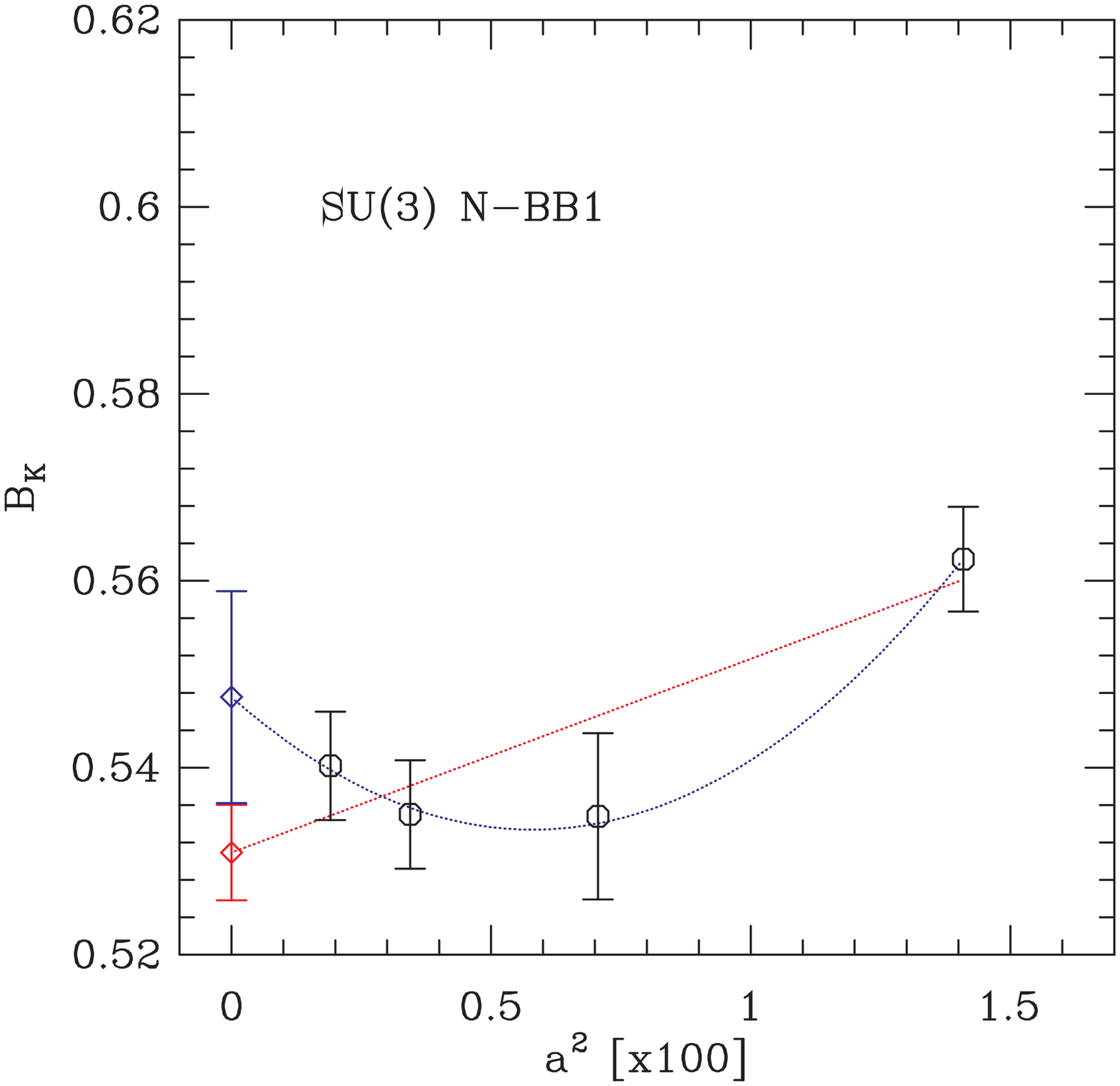}
\includegraphics[width=0.49\textwidth]{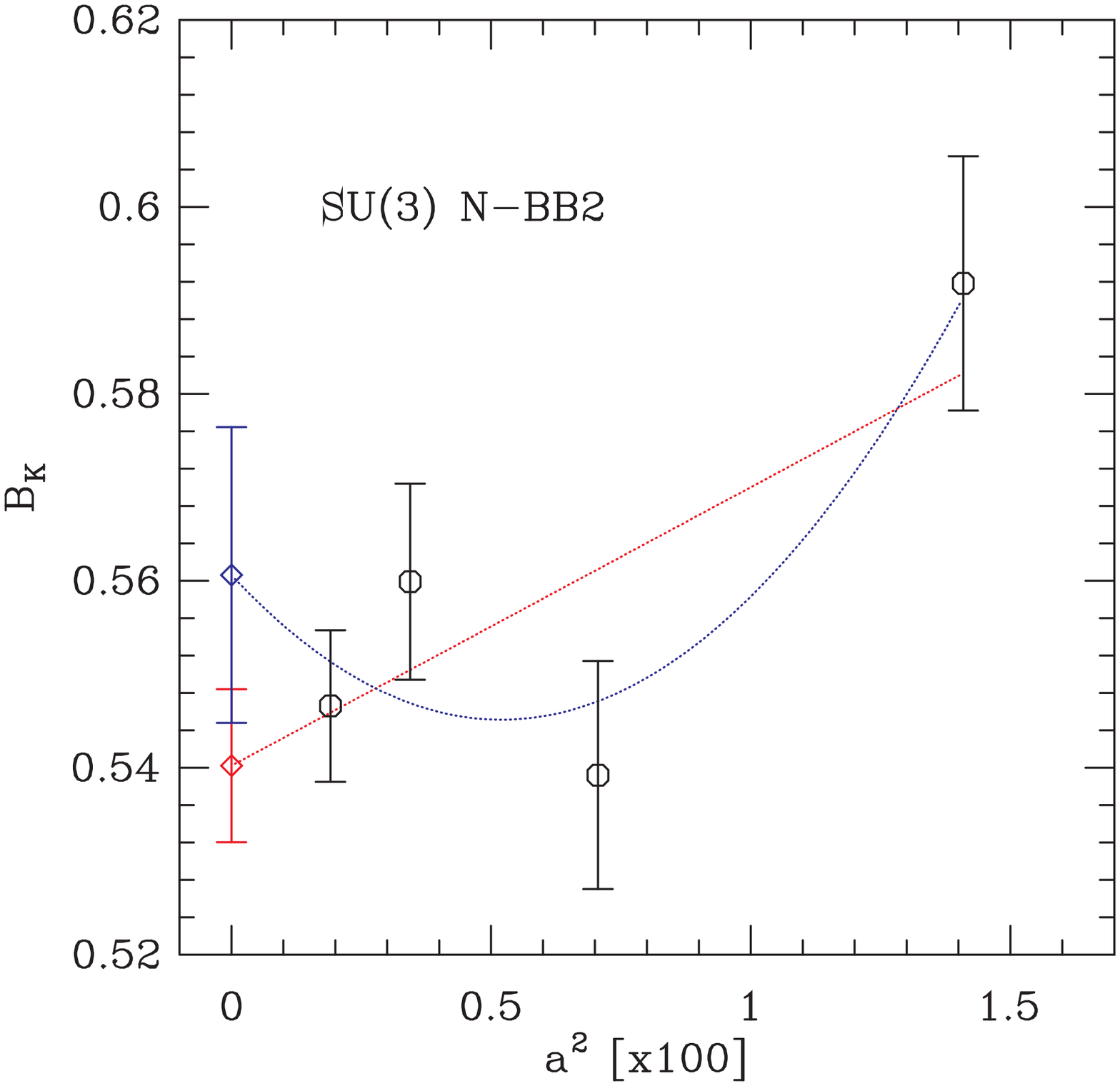}
\caption{ $B_K(\text{NDR}, \mu=2\text{ GeV})$ as a function of $a^2$ 
(in fm $\times 100$) for the N-BB1 fit (left) and the N-BB2 fit (right), 
showing linear and quadratic fits to $a=0$.
}
\label{fig:bk:a^2}
\end{figure}

The expected dependence on $a$ is due to both discretization errors
of the form $a^2 \alpha_s^n$, with $n=0,1,2\dots$,
and truncation errors starting at order $\alpha_s^2$,
with $\alpha_s$ evaluated at a scale $\sim 1/a$~\cite{ref:wlee-2010-1}.
We assume that the former dominate, with $n=0$,
and correct this assumption by adding in appropriate systematic errors.
Thus we extrapolate using both linear and quadratic dependence on $a^2$,
as shown in Fig.~\ref{fig:bk:a^2}.

The data are consistent with both fit forms, although the quadratic
fits are somewhat preferred. The parameters of the quadratic fit are,
however, implausible. We expect the relative size of the
quadratic and linear terms to be $\sim (a\Lambda)^2$, which, 
with $\Lambda=500\;$MeV and for the coarse lattices is
$\sim 0.09$. Instead, in the quadratic fits the linear and quadratic
terms are comparable on the coarse lattices.
Thus we use the linear fits for our central values, and take the
difference with the quadratic fit as the systematic error due
to continuum extrapolation.

The truncation error we estimate separately by assuming the
missing terms in the matching factor have size
$1 \times \alpha_s(1/a)^2$, as explained further in Ref.~\cite{ref:wlee-2010-1}.

\section{Error Budget and Conclusions}
%
%
%
%-----------------------------------
% error budget of the SU(3) fitting
%-----------------------------------
\begin{table}[h]
\centering
\begin{tabular}{ l | l l r }
\hline \hline
cause & error (\%) & memo  & status \\
\hline
statistics       & 1.0  & N-BB1 fit & update \\
matching factor  & 4.4   & $\Delta B_K^{(2)}$ (S1) & update \\
discretization   & 3.1   & diff.~of linear and quadratic extrap & update\\
fitting (1)      & 0.36  & diff.~of N-BB1 and N-B1 (C3) & \cite{ref:wlee-2010-1}\\
fitting (2)      & 5.3   & diff.~of N-BB1 and N-BB2 (C3)& \cite{ref:wlee-2010-1} \\
$a m_l$ extrap   & 1.0   & diff.~of (C3) and linear extrap & \cite{ref:wlee-2010-1}\\
$a m_s$ extrap   & 0.5   & constant vs. linear extrap & \cite{ref:wlee-2010-1}\\
finite volume    & 2.3   & diff.~of $20^3$ (C3) and $28^3$ (C3-2)& \cite{ref:wlee-2010-1}\\
scale $r_1$      & 0.12  & uncertainty in $r_1$ & \cite{ref:wlee-2010-1}\\
\hline \hline
\end{tabular}
\caption{Error budget for $B_K$ obtained using SU(3) SChPT fitting.
  \label{tab:su3-err-budget}}
\end{table}
In Table \ref{tab:su3-err-budget}, we collect our estimates of
all sources of error, noting which results have been updated from
our article~\cite{ref:wlee-2010-1}. We refer to that reference for
details of the unchanged estimates.

Compared to our result based on 3 lattice spacings~\cite{ref:wlee-2010-1}, 
the statistical error has decreased (from 1.4\%), as
has the error due to the matching factor (from 5.5\%).
Both decreases are due to our addition of a fourth lattice spacing.
On the other hand, our estimate of the discretization error has
increased (from 2.2\%). This is because we now use a different,
more conservative method, namely the difference between linear and
quadratic fits, rather than the difference between the result on
ensemble S1 and the continuum value.

The largest error is now the ``fitting (2)'' error, which is our estimate
of the uncertainty related to the different choices of priors in the
Bayesian fits. We obtain this error from the difference between the
results of the N-BB1 and N-BB2 fits on the coarse lattices.
We could use the difference between these two fits after continuum
extrapolation, which would more than halve the error, but we are not
sufficiently confident in the continuum extrapolation to do so.
In particular, the difference between these two fits remains substantial
on the S1 ensemble, whereas one would expect it to decrease compared to
the C3 ensemble, since lattice artifacts are substantially smaller.

Combining the systematic errors in quadrature, our current result for
$B_K$ using SU(3) SChPT fitting is
\begin{equation}
\begin{array}{l l}
  B_K(\text{NDR}, \mu = 2 \text{ GeV}) & = 0.5309 \pm 0.0051 \pm 0.0424\,,
  \\
  \hat{B}_K = B_K(\text{RGI}) & = 0.7270 \pm 0.070 \pm 0.0580\,,
\end{array}
\end{equation}
where the first error is statistical and the second is systematic.
This updates our result $B_K(2\;{\rm GeV}=0.524\pm0.007\pm 0.044$
given in Ref.~\cite{ref:wlee-2010-1}.
Compared with the SU(2) SChPT analysis~\cite{ref:wlee-2010-2}, 
the statistical error is smaller here
but the systematic error is significantly larger.
Overall the SU(3) result has a larger error (8\% vs. 5\%).
Given the less straightforward fitting, the concern with
convergence of SU(3) SChPT, and the larger error, we use
the SU(3) analysis as a cross-check on our result from
the preferred SU(2) analysis. The results are consistent.

\section{Acknowledgments}
C.~Jung is supported by the US DOE under contract DE-AC02-98CH10886.
The research of W.~Lee is supported by the Creative Research
Initiatives Program (3348-20090015) of the NRF grant funded by the
Korean government (MEST). 
The work of S.~Sharpe is supported in part by the US DOE grant
no.~DE-FG02-96ER40956.
Computations were carried out in part on QCDOC computing facilities of
the USQCD Collaboration at Brookhaven National Lab. The USQCD
Collaboration are funded by the Office of Science of the
U.S. Department of Energy.

\end{document}